\documentclass[11pt]{article}
\usepackage{amsmath}
\usepackage{amssymb}
\usepackage{epsfig}
\newlength{\bredde}
\def\slash#1{\settowidth{\bredde}{$#1$}\ifmmode\,\raisebox{.15ex}{/}
\hspace*{-\bredde} #1\else$\,\raisebox{.15ex}{/}\hspace*{-\bredde} #1$\fi}
\newcommand{\be}{\begin{equation}}
\newcommand{\ee}{\end{equation}}
\newcommand{\nn}{\nonumber}

\newcommand{\la}{\lambda}

\newcommand{\sig}{\sigma}

\newcommand{\eins}{\leavevmode\hbox{\small1\kern-3.8pt\normalsize1}}

\newcommand{\e}{\mbox{e}}
\newcommand{\sign}{\mbox{sign}}

\newcommand{\erfc}{\mbox{erfc}}
\newcommand{\Ai}{\mbox{Ai}}

\newcommand{\mcZ}{\mathcal{Z}}

\newcommand{\dsq}{d^{\hspace{0.5pt}2}}

\newcommand{\sect}[1]{\setcounter{equation}{0}\section{#1}}

\DeclareMathOperator*{\Tr}{Tr}

\DeclareMathOperator*{\Pf}{Pf}

\def\gb{\gamma_\beta}

\begin{document}
\topmargin -1.4cm
\oddsidemargin -0.8cm
\evensidemargin -0.8cm
\title{\Large
{{\bf 
Universality Conjecture for all Airy, Sine and Bessel Kernels in the 
Complex Plane 
}}}

\vspace{1.5cm}
\author{~\\{\sc G.~Akemann}$^1$ and {\sc M.~J.~Phillips}$^{2}$
\\~\\$^1$Department of Physics,
Bielefeld University,\\
Postfach 100131,
D-33501 Bielefeld, Germany
\\~\\
$^2$School of Mathematical Sciences, Queen Mary, University of London, 
\\ London E1 4NS, United Kingdom
}

\date{}

\maketitle
\vfill
\begin{abstract}
We address the question of how the celebrated universality of local
correlations for the real eigenvalues of Hermitian 
random matrices of size $N\times N$ can be extended to complex
eigenvalues in the case of random matrices without symmetry.
Depending on the location in the spectrum, 
particular large-$N$ limits (the so-called weakly non-Hermitian limits) lead to
one-parameter deformations of the Airy, sine and Bessel kernels into
the complex plane. 
This makes their universality highly suggestive for all symmetry classes.
We compare all the known limiting real kernels 
and their deformations into the complex
plane for all three 
Dyson indices
$\beta=1,2,4$, corresponding to real, complex and
quaternion real matrix elements.
This includes new results for Airy kernels in the complex plane for
$\beta=1,4$. For the Gaussian ensembles of elliptic Ginibre and
non-Hermitian Wishart matrices we
give all kernels for finite $N$, built from 
orthogonal and skew-orthogonal polynomials in the complex plane.
Finally we comment on how much is known to date regarding the
universality of these kernels in the complex plane, and discuss some
open problems.

\end{abstract}
\vfill

\thispagestyle{empty}
\newpage

\renewcommand{\thefootnote}{\arabic{footnote}}
\setcounter{footnote}{0}

\sect{Introduction}\label{intro}

The topic of universality in Hermitian Random Matrix Theory (RMT)
has attracted a lot of attention in the mathematics community
recently, particularly in the context of matrices with elements that are
independent random variables, as reviewed in \cite{EYTV}. 
The question that one tries to answer is this: Under what conditions are the
statistics of eigenvalues of $N\times N$ matrices with independent Gaussian
variables the same (for large matrices) as for 
more general RMT
where matrix elements may become coupled? This has been answered under very
general assumptions, and we refer to some recent reviews on invariant
\cite{Arno,Deift} and non-invariant \cite{EYTV} ensembles.

In this short note we would like to advocate the idea that
non-Hermitian RMT with eigenvalues in the complex plane also warrants
the investigation of universality. Apart from the interest in its own
right, these models have important applications in physics and other
sciences (see e.g.\ \cite{ABD}).
We will focus here on RMT that is close to Hermitian, a regime which is
particularly important for applications in 
quantum chaotic scattering (see \cite{FyoSom03} for a review) and
Quantum Chromodynamics (QCD), for example. In the latter case,
the non-Hermiticity may arise from describing the effect of quark
chemical potential (as reviewed
in \cite{Jac}), or from finite lattice spacing effects
of the Wilson-Dirac operator (see \cite{ADSV} as well as
\cite{Mario} for the solution of this non-Hermitian RMT).

Being a system of $N$ coupled eigenvalues,
Hermitian RMT already offers a rich variety of large-$N$ limits, where
one has to distinguish the bulk and (soft) edge of the spectrum for
Wigner-Dyson (WD) ensembles, and in addition the origin (hard edge) 
for Wishart-Laguerre (WL, or chiral) RMT.
Not surprisingly complex eigenvalues
offer even more possibilities. The 
limit we will investigate
is known as the
weakly non-Hermitian regime; it
{\it connects} Hermitian and (strongly) non-Hermitian RMT, and was
first introduced  
in \cite{FKS} in the bulk of the spectrum. For strong non-Hermiticity --
which includes the well-known 
circular law and the corresponding 
universality results --
we refer to \cite{BHJS} and references therein,
although the picture here is also far from being
complete. 

In the next section
we give a brief list of the six non-Hermitian WD and
WL ensembles, and indicate where they were first solved in the weak limit.
There are three principal reasons why we believe that universality may hold.
First, in some cases two different Gaussian RMT both give the same answers. 
Second, there are heuristic arguments
available for i.i.d.\ matrix elements using supersymmetry \cite{FKS98},
as well as for invariant non-Gaussian 
ensembles using large-$N$ factorisation and
orthogonal polynomials (OP) \cite{Ake02}. 
Third, the resulting limiting kernels of
(skew-) OP look very similar to the corresponding kernels of real eigenvalues,
being merely one-parameter deformations of them. 
One of the main goals
of this paper is to illustrate this fact. For this purpose we give a
complete list of all the known Airy, sine, and Bessel kernels for
real eigenvalues,
side-by-side with their deformed kernels in the complex plane, where some of
our results are new.

\sect{Random matrices and their limiting kernels}\label{results}

In this section we briefly introduce the Gaussian random matrix
ensembles that we consider,
and give a list of the limiting kernels they lead to, for
both real and complex eigenvalues. For simplicity we have
restricted ourselves to Gaussian ensembles in the Hermitian cases, in
order to highlight the parallels to their non-Hermitian counterparts.

We begin with the classical WD
and Ginibre ensembles in \textsection\ref{SecGaussianEnsembles},
displaying Airy (\textsection\ref{secLimitingAiry}) 
and sine (\textsection\ref{secLimitingSine}) behaviour
at the (soft) edge and in the bulk of the spectrum respectively, as
well as their deformations. We then introduce the
WL ensembles and their non-Hermitian
counterparts in \textsection\ref{WL}, in order to access the Bessel
behaviour (\textsection\ref{secLimitingBessel}) at the origin (or hard edge).
The corresponding orthogonal and skew-orthogonal
Hermite and Laguerre
polynomials are given in Appendix \ref{SOP}, and
precise statements of the limits that lead to the microscopic kernels can
be found in Appendix \ref{evalscalings}.

\subsection{Gaussian ensembles with eigenvalues 
on ${\mathbb R}$ and ${\mathbb C}$}\label{SecGaussianEnsembles}

The three classical Gaussian Wigner-Dyson ensembles 
(the GOE, GUE and GSE) are defined as \cite{Mehta}
\be
\mcZ_{N}^{\text{G$\beta$E}}= \int dH \exp[-\beta \Tr H^2/4]=c_{N,\beta}
\prod_{j=1}^N \int_{{\mathbb R}}
dx_j\,w_{\beta}(x_j)\ |\Delta_N(\{x\})|^\beta\ .
\label{GbE}
\ee
The random matrix elements $H_{kl}$ are real, complex, or quaternion
real numbers for $\beta=1,2,4$ respectively, with the condition that
the $N\times N$ matrix $H$
($N$ is taken to be even for simplicity)
is real symmetric, complex Hermitian or complex Hermitian and self-dual for
$\beta=1,2,4$. 
In the first equation we integrate over all independent
matrix elements denoted by $dH$. The Gaussian weight completely
factorises and thus the independent elements are normal random variables;
for $\beta=1$, for example, the real elements are distributed
${\cal N}(0,1)$ for off-diagonal elements, and ${\cal
  N}(0,\sqrt{2})$ for diagonal elements.

In the second step above, we diagonalised the matrix
$H=U$diag$(x_1,\ldots,x_N)U^{-1}$ where $U$ is an orthogonal, unitary
or unitary-symplectic matrix for $\beta=1,2,4$. The integral over the
latter factorises and leads to the known constants $c_{N,\beta}$. We
obtain a Gaussian weight $w_\beta(x)$
and the Vandermonde determinant $\Delta_N(\{x\})$
from the Jacobian of the diagonalisation,
\be
w_\beta(x)=\exp[-\beta x^2/4]\ , \ \ \Delta_N(\{x\})
=\prod_{1\leq l<k\leq N}(x_k-x_l)\ .
\label{wGauss}
\ee
The integrand on the right-hand side
of eq.\ (\ref{GbE}) times $c_{N,\beta}/\mcZ_{N}^{\text{G$\beta$E}}$
defines the normalised joint probability distribution function (jpdf) of all
eigenvalues. The $k$-point correlation function $R_k^\beta$, which is
proportional to the jpdf integrated over $N-k$ eigenvalues, can be
expressed through a single kernel $K_{N}^{\beta=2}$
of orthogonal polynomials (OP) for $\beta=2$,
or through a $2\times2$ matrix-valued kernel
involving skew-OP for $\beta=1,4$:
\begin{align}
R_{k}^{\beta=2}(x_1,\ldots,x_k)
&=\det_{i,j=1,\ldots,k}[K_{N}^{\beta=2}(x_i,x_j)]
\ ,\nn\\
R_{k}^{\beta=1,4}(x_1,\ldots,x_k)&=\Pf_{i,j=1,\ldots,k}
\left[\left(
\begin{array}{cc}
K_{N}^{\beta=1,4}(x_i,x_j)
&-G_{N}^{\beta=1,4}(x_i,x_j)\\
G_{N}^{\beta=1,4}(x_j,x_i)
&-W_{N}^{\beta=1,4}(x_i,x_j)\\
\end{array}
\right)
\right].
\label{Rkreal}
\end{align}
The matrix kernel elements $K_N$ and $W_N$ are not independent of $G_N$ but
are related by differentiation and integration respectively. These
relations will be given later for the limiting kernels.

The three parameter-dependent 
Ginibre (i.e.\ elliptic or Ginibre-Girko) ensembles,
denoted by GinOE, GinUE, and GinSE,
can be written as
\be
\mcZ_{N}^{\text{Gin$\beta$E}}(\tau)= \int dJ\exp\left[\frac{-\gb}{1-\tau^2}
\Tr\Big(
JJ^\dag-\frac{\tau}{2}(J^2+J^{\dag\,2})\Big)\right]=
\int dH_1\,dH_2 \exp\left[-\frac{\gb\Tr H_1^2}{1+\tau}- 
\frac{\gb\Tr H_2^2}{1-\tau}\right]\!,
\ \ \ 
\label{GinbE}
\ee
with $\tau\in[0,1)$. We use the parametrisation of \cite{BHJS}, with
  $\gamma_{\beta=2}=1$ and $\gamma_{\beta=1,4}=\tfrac12$.
The matrix elements of $J$ are of the same types as for $H$
  for all three values of 
$\beta$, but without any further symmetry constraint.
Decomposing $J=H_1+iH_2$ into its Hermitian and
  anti-Hermitian parts, these ensembles can be viewed as 
  Gaussian two-matrix models. 
For $\tau=0$ (maximal non-Hermiticity) 
the distribution for all matrix elements again
factorises. In the opposite, i.e.\ Hermitian, limit,
the Ginibre ensembles become the Wigner-Dyson cases.
The jpdf of complex (and real) eigenvalues can be computed
by transforming $J$ into the following form, $J=U(Z+T)U^{-1}$. For
$\beta=2$ this is the Schur decomposition, with $Z=$diag$(z_1,\ldots,z_N)$
containing the complex eigenvalues, and $T$ being upper
triangular\footnote{
The resulting jpdf of complex eigenvalues for normal matrices with
$T\equiv 0$ at $\beta=2$  
is the same.}
\be
\mcZ_{N}^{\text{GinUE}}(\tau)=c_{N,\mathbb C}^{\beta=2}
\prod_{j=1}^N\int_{\mathbb C}\dsq z_j\,w_{\beta=2}^{\mathbb C}(z_j)\ 
|\Delta_N(\{z\})|^2,
\ \ w_{\beta=2}^{\mathbb C}(z)=\exp\left[\frac{-1}{1-\tau^2}
\Big(
|z|^2-\frac{\tau}{2}(z^2+z^{*\,2})\Big)\right]. \ \ 
\label{GinUE}
\ee
For $\beta=1,4$ we follow \cite{BHJS} where the two ensembles have
been cast into a unifying framework. For simplicity we choose $N$ to
be even.
Here the matrix $Z$ can be chosen to be $2\times2$ block
diagonal and $T$ to be upper block triangular. The calculation of the
jpdf reduces to a $2\times2$ calculation, yielding
\be
\mcZ_{N}^{\text{GinO/SE}}(\tau) = c_{N,\mathbb C}^{\beta=1,4}
\prod_{j=1}^N\int_{\mathbb  C} \dsq z_j\,
\prod_{k=1}^{N/2} {\cal F}_{\beta=1,4}^{\mathbb C}(z_{2k-1},z_{2k})\, 
\Delta_N(\{z\})\ ,
\label{GinSOE}
\ee
where we have introduced an anti-symmetric bivariate weight function.
For $\beta=1$, this is given by
\begin{align}
{\cal F}^{\mathbb C}_{\beta=1}(z_1,z_2) &= w_{\beta=1}^{\mathbb C}(z_1)
w_{\beta=1}^{\mathbb C}(z_2)\Big( 2i\delta^{2}(z_1-z_2^*)
\sign(y_1)
+\delta^{1}(y_1)\delta^{1}(y_2)\sign(x_2-x_1)
\Big), \label{Fb1} \\
\Big(w_{\beta=1}^{\mathbb C}(z)\Big)^2
&= \erfc\left(\frac{|z-z^*|}{\sqrt{2(1-\tau^2)}}\right)
\exp\left[\frac{-1}{2(1+\tau)}(z^2+z^{*\,2})\right], \nn
\end{align}
and for $\beta=4$ by
\be
{\cal F}^{\mathbb C}_{\beta=4}(z_1,z_2) = w_{\beta=4}^{\mathbb C}(z_1)
w_{\beta=4}^{\mathbb C}(z_2) \, (z_1-z_2) \, \delta(z_1-z_2^*), \qquad
\Big(w_{\beta=4}^{\mathbb C}(z)\Big)^2 = 
w_{\beta=2}^{\mathbb C}(z)\ . \label{Fb4}
\ee
For $\beta=1$, it should be noted that the integrand in
eq.\ (\ref{GinSOE}) is not always positive, 
and so a symmetrisation must be applied when determining
the correlation functions below\footnote{It is, however, possible to write the
partition function ${\cal Z}_N^{\text{GinOE}}$ as an integral over a
true (i.e.\ positive) jpdf, by, 
for example, appropriately ordering the eigenvalues; however, such a
representation is technically more difficult to work with.}.
For $\beta=4$, the parameter $N$ in eq.\ (\ref{GinSOE}) should -- in
our convention -- be taken to be the size of the complex-valued matrix
that is equivalent to the original quaternion real matrix.

The correlation functions can be written in a similar form
as for the real eigenvalues
\begin{align}
R_{k,\mathbb C}^{\beta=2}(z_1,\ldots,z_k)
 &= \det_{i,j=1,\ldots,k}[K_{N,\mathbb C}^{\beta=2}(z_i,z_j^*)]
\ ,\nn\\
R_{k,\mathbb C}^{\beta=1,4}(z_1,\ldots,z_k) &= \Pf_{i,j=1,\ldots,k}
\left[\left(
\begin{array}{cc}
K_{N,{\mathbb C}}^{\beta=1,4}(z_i,z_j)
&-G_{N,\mathbb C}^{\beta=1,4}(z_i,z_j)\\
G_{N,\mathbb C}^{\beta=1,4}(z_j,z_i)
&-W_{N,\mathbb C}^{\beta=1,4}(z_i,z_j)\\
\end{array}
\right)\right],
\label{Rkcomp}
\end{align}
where the elements of the matrix kernels are related through
\begin{align}
G_{N,\mathbb C}^{\beta=1,4}(z_i,z_j) &= -\int_{\mathbb C} \dsq z
K_{N,{\mathbb C}}^{\beta=1,4}(z_i,z){\cal F}^{\mathbb
  C}_{\beta=1,4}(z,z_j)\ ,
\nn\\
W_{N,\mathbb C}^{\beta=1,4}(z_i,z_j) &= \int_{\mathbb C^2} \dsq z \, \dsq z' \,
{\cal F}^{\mathbb C}_{\beta=1,4}(z_i,z) K_{N,{\mathbb C}}^{\beta=1,4}(z,z'){\cal
  F}^{\mathbb C}_{\beta=1,4}(z',z_j) - {\cal F}^{\mathbb
  C}_{\beta=1,4}(z_i,z_j)\ . 
\label{GKWrel}
\end{align}
The kernels $K_{N,{\mathbb C}}^{\beta}(z,z')$ are given explicitly in
Appendix \ref{SOP}. 

For $\beta=1$, we can write
\be
G_{N,\mathbb C}^{\beta=1}(z_1,z_2) =
\delta^{1}(y_2)
G_{N.\mathbb C,\text{real}}^{\beta=1}(x_1,x_2)
+G_{N,\mathbb
  C,\text{com}}^{\beta=1}(z_1,z_2),
\ee
whereas for $\beta=4$ eq.\ (\ref{Fb4}) implies the following relations:
\begin{align}
G_{N,\mathbb  C}^{\beta=4}(z_1,z_2) &= (z_2-z_2^*)
w_{\beta=2}^{\mathbb C}(z_2) K_{N,{\mathbb C}}^{\beta=4}(z_1,z_2^*)\ ,
\nn\\
W_{N,\mathbb C}^{\beta=4}(z_1,z_2) &= -(z_1-z_1^*)(z_2-z_2^*)
w_{\beta=2}^{\mathbb C}(z_1) w_{\beta=2}^{\mathbb C}(z_2) 
K_{N,{\mathbb C}}^{\beta=4}(z_1^*,z_2^*)
\ ,
\label{GKWrelb4}
\end{align}
(where we dropped the `contact' term in the final expression).
For this reason, for $\beta=4$ we will only give 
one of the matrix kernel elements
in the following. 

Note that $\beta=1$ is special as the eigenvalues of a real asymmetric
matrix are either real or come in complex conjugate pairs.
Therefore we will have to distinguish kernels (and $k$-point densities)
of real, complex or mixed arguments.

In order to specify the limiting kernels we first need the behaviour
of the mean (or macroscopic) spectral density. At large $N$,
and for all three values of $\beta$,
the (real) eigenvalues in the Hermitian cases are predominantly
concentrated within the Wigner 
semi-circle $\rho_{\text{sc}}(x)=(2\pi N)^{-1}\sqrt{4N-x^2}$
on $[-2\sqrt{N},2\sqrt{N}]$, whereas in the non-Hermitian case, the
complex eigenvalues lie 
mostly within an ellipse with half-axes of lengths $(1+\tau)\sqrt{N}$ and
$(1-\tau)\sqrt{N}$, with constant
density $\rho_{\text{el}}(z)=(N\pi(1-\tau^2))^{-1}$. 
Depending on where
(and how) we magnify the spectrum locally, we obtain different asymptotic
Airy or sine kernels for each $\beta=1,2,4$. In the following we will
give all of the known real kernels, see e.g.\ \cite{Arno} for a
complete list and references, together with their deformations into the
complex plane.
For the Bessel kernels which will be shown later we
need to consider different matrix ensembles, see \textsection\ref{WL} below.

\subsection{Limiting Airy kernels on ${\mathbb R}$ and ${\mathbb
    C}$}\label{secLimitingAiry} 

When appropriately zooming into the ``square root'' edge of the
semi-circle,
the three well-known Airy kernels (matrix-valued for $\beta=1,4$)
are obtained for real eigenvalues. 
For complex eigenvalues we have to consider the vicinity of the eigenvalues
on a thin ellipse which have the largest real parts, and where the
weakly non-Hermitian 
limit introduced in \cite{Bender} is defined such that
\be
\sigma=N^{\frac16} \,\sqrt{1-\tau}
\label{sig-Ai}
\ee
remains fixed (see Appendix \ref{evalscalings} for the precise details of
the scaling of the eigenvalues). This leads to one-parameter
deformations of the Airy kernels
in the complex plane. Whilst the result for $\beta=2$
is already known \cite{Bender,ABe}, our results for $\beta=1,4$ stated below are
new \cite{APh}.
\begin{align}
\underline{\beta=2:} &&
K_{\text{Ai}}^{\beta=2}(x_1,x_2) &=
\frac{\Ai(x_1)\Ai'(x_2)-\Ai'(x_1)\Ai(x_2)}{x_1-x_2} 
= \int_0^\infty dt \, \Ai(x_1+t)\Ai(x_2+t)\ ,
\label{AiKb2}\\
&&\nn\\
&& K_{\text{Ai},\mathbb C}^{\beta=2}(z_1,z_2) &= \frac{1}{\sig\sqrt{\pi}}\,
\e^{-\frac{y_1^2+y_2^2}{2\sig^2}+\frac{\sig^6}{6}
+\frac{\sig^2(z_1+z_2)}{2}}
\int_0^\infty dt\ \e^{\sig^2 t}
\Ai\Big(z_1+t+\frac{\sig^4}{4}\Big)\Ai\Big(z_2+t+\frac{\sig^4}{4}\Big).
\label{AiKb2C}
\end{align}
In the Hermitian limit $\sig\to0$ we obtain
$K_{\text{Ai},\mathbb C}^{\beta=2}(z_1,z_2) \rightarrow
\sqrt{\delta^{1}(y_1)\delta^{1}(y_2)}
K_{\text{Ai}}^{\beta=2}(x_1,x_2)$,
with the factor in front of the integral in eq.\ (\ref{AiKb2C}) 
projecting the imaginary parts of the eigenvalues to zero.
For the integral itself -- which is obtained from the limit of the sum
of the OP on $\mathbb C$ given in eq.\ (\ref{Kb2C}) -- the deformation
in $\sig$ is very smooth. 
The same deformed Airy kernel can be obtained from the corresponding WL ensemble
eq.\ (\ref{CLbE}) \cite{ABe} with kernel eq.\ (\ref{Kb2Cch}), and is
thus universal.

\begin{align}
\underline{\beta=4:} && G_{\text{Ai}}^{\beta=4}(x_1,x_2)&=
- \tfrac12 K_{\text{Ai}}^{\beta=2}(x_1,x_2) + \tfrac14
\Ai(x_1)\int_{x_2}^\infty dt \, \Ai(t),\ 
K_{\text{Ai}}^{\beta=4}(x_1,x_2)=\frac{\partial}{\partial x_2}
G_{\text{Ai}}^{\beta=4}(x_1,x_2),
\nn\\
&& W_{\text{Ai}}^{\beta=4}(x_1,x_2) 
&= -\int_{x_1}^\infty ds\ G_{\text{Ai}}^{\beta=4}(s,x_2)\nn\\
&& &= - \tfrac14\int_0^\infty ds\int_0^s dt\Big( 
\Ai(x_2+t)\Ai(x_1+s)- \Ai(x_2+s)\Ai(x_1+t) \Big),
\label{AiKb4}\\
&& G_{\text{Ai},\mathbb C}^{\beta=4}(z_1,z_2)
&=\frac{iy_2}{4\sig^3\sqrt{\pi}}
\e^{-\frac{y_1^2+y_2^2}{2\sig^2}+\frac{\sig^6}{6}
+\frac{\sig^2(z_1+z_2^*)}{2}} \label{AiKb4C}
\\
&& & \ \ \ \  \times\int_0^\infty ds \int_0^s dt \, e^{\frac12\sig^2 (s+t)}
\left(\Ai\Big(z_2^*+s+\frac{\sig^4}{4}\Big)
\Ai\Big(z_1+t+\frac{\sig^4}{4}\Big)- (z_1\leftrightarrow z_2^*)\right)\!.\nn
\end{align}
The integral in eq.\ (\ref{AiKb4C}) which is also present in the
other two kernel elements, see eq.\ (\ref{GKWrelb4}), clearly reduces to that in
eq.\ (\ref{AiKb4}) in the Hermitian limit, whereas the pre-factors
provide the appropriate Dirac delta functions.
When analysing the Hermitian limit in detail, the real kernel elements
$G_{\text{Ai}}^{\beta=4}$ and $K_{\text{Ai}}^{\beta=4}$ follow from a
Taylor expansion of $W_{\text{Ai},\mathbb C}^{\beta=4}$, see 
\cite{ABa} for a discussion of the analogous Hermitian limit of
the Bessel kernel. 
\begin{align}
\underline{\beta=1:}
&& G_{\text{Ai}}^{\beta=1}(x_1,x_2)&
= - \int_0^\infty dt \,  \Ai(x_1+t)\Ai(x_2+t) - \tfrac12 \Ai(x_1)
\left( 1-\int_{x_2}^\infty dt \, \Ai(t)\right)\ ,\nn\\ 
&& K_{\text{Ai}}^{\beta=1}(x_1,x_2)&=\frac{\partial}{\partial x_2}
G_{\text{Ai}}^{\beta=1}(x_1,x_2)\ ,\ \nn\\
&& W_{\text{Ai}}^{\beta=1}(x_1,x_2)&=-\int_{x_1}^\infty ds\ 
G_{\text{Ai}}^{\beta=1}(s,x_2)
-\tfrac12\int_{x_1}^{x_2}dt\ \Ai(t)
+\tfrac12\int_{x_1}^\infty ds\ \Ai(s)\int_{x_2}^\infty dt\ \Ai(t)\nn\\
&& & \qquad -\tfrac12\ \sign(x_1-x_2)\ ,
\label{AiKb1} \\
&& G_{\text{Ai},\mathbb C,\text{real}}^{\beta=1}(x_1,x_2)&=
- \e^{\frac{\sig^6}{6}+\frac{\sig^2(x_1+x_2)}{2}}\int_0^\infty dt\ \e^{\sig^2 t}
\Ai\Big(x_1+t+\frac{\sig^4}{4}\Big)\Ai\Big(x_2+t+\frac{\sig^4}{4}\Big)\nn\\
&& & \qquad -\tfrac12\ \e^{\frac{\sig^6}{12}+\frac{\sig^2
    x_1}{2}}\Ai\Big(x_1+\frac{\sig^4}{4}\Big) 
\left(1-\e^{\frac{\sig^6}{12}}\int_{x_2}^\infty dt\ \e^{\sig^2 t/2}
\Ai\Big(t+\frac{\sig^4}{4}\Big) 
\right)\ ,\nn\\
&& G_{\text{Ai},\mathbb
  C,\text{com}}^{\beta=1}(z_1,z_2) &=  - \, \frac{i}{2\sig^2} \,
\sign(y_2)(z_1-z_2^*) \, 
\e^{\frac{\sig^6}{6}+\frac{\sig^2(x_1+x_2)}{2}}
\sqrt{\erfc\left(\frac{|y_1|}{\sig}\right)  
  \erfc\left(\frac{|y_2|}{\sig}\right)}\nn\\ 
&& & \qquad \times\int_0^\infty dt \left( \e^{\sig^2
  t}-1\right)\Ai\Big(z_1+t+\frac{\sig^4}{4}\Big)
\Ai\Big(z_2^*+t+\frac{\sig^4}{4}\Big)\ ,\nn\\ 
&& K_{\text{Ai},{\mathbb C}}^{\beta=1}(z_1,z_2) &= \frac{i}{2} \,
\sign(y_2)
G_{\text{Ai},\mathbb C,\text{com}}^{\beta=1}(z_1,z_2^*)\ ,\nn\\
&& W_{\text{Ai},\mathbb C}^{\beta=1}(z_1,z_2)&= 
\Big( -2A(x_1,x_2)+B(x_1)B(x_2)+B(x_2)-B(x_1)
\Big)\delta^1(y_1)\delta^1(y_2) \nn\\ 
&& & \qquad {} + 2i \, \Big(\sign(y_2)
G_{\text{Ai},\mathbb  C,\text{real}}^{\beta=1}(z_2^*,z_1)- 
\sign(y_1)G_{\text{Ai},\mathbb C,\text{real}}^{\beta=1}(z_1^*,z_2)\Big)\nn\\
&& & \qquad -2i \, \sign(y_1)G_{\text{Ai},\mathbb
  C,\text{com}}^{\beta=1}(z_1^*,z_2) \nn \\
&& & \qquad -2i\delta^2(z_1-z_2^*)
\sign(y_1)-\delta^1(y_1)\delta^1(y_2)  
\sign(x_2-x_1)\ ,\nn\\
&& A(x_1,x_2) &= \e^{\frac{\sigma^6}{6}+\frac{\sigma^2(x_1+x_2)}{2}}
\int_0^{\infty} ds \int_0^s dt \, \e^{\frac{1}{2}\sigma^2(s+t)} \Ai
\Big(x_1+s+\frac{\sig^4}{4}\Big) \Ai \Big(x_2+t+\frac{\sig^4}{4}\Big),
\nn \\ 
&& B(x) &= \e^{\frac{\sigma^6}{12}+\frac{\sigma^2 x}{2}}
\int_0^{\infty} dt \, \e^{\frac{1}{2}\sig^2t} \Ai
\Big(x+t+\frac{\sig^4}{4}\Big). 
\label{AiKb1C}
\end{align}
Clearly it holds that 
$G_{\text{Ai},\mathbb C,\text{real}}^{\beta=1}(x_1,x_2) \rightarrow
G_{\text{Ai}}^{\beta=1}(x_1,x_2)$ as $\sigma\rightarrow 0$, whereas
the complex part vanishes  
in this Hermitian limit 
$G_{\text{Ai},\mathbb C,\text{com}}^{\beta=1}(z_1,z_2) \rightarrow 0$.
We have also explicitly verified the corresponding limits for
$K_{\text{Ai},{\mathbb C}}^{\beta=1}(z_1,z_2)$ 
and $W_{\text{Ai},\mathbb C}^{\beta=1}(z_1,z_2)$.

\subsection{Limiting sine kernels on ${\mathbb R}$ and ${\mathbb
    C}$}\label{secLimitingSine} 

For real eigenvalues the sine kernels are obtained by zooming into
the bulk of the spectrum, sufficiently far away from the edges.
The weakly non-Hermitian limit of the complex eigenvalues introduced in
\cite{FKS} is taken such that  
\be
\sig=N^{1/2}\sqrt{1-\tau}
\label{sig-sine}
\ee
remains finite (see Appendix \ref{evalscalings} for further details).
In this limit the macroscopic support of
the spectral density on an ellipse 
shrinks to the semi-circle distribution on the real axis, whereas
microscopically 
we still have correlations of the eigenvalues in the complex plane.

The list of the known one-parameter deformations of
the sine kernels is as follows: 
\begin{align}
\underline{\beta=2:}
&&
K_{\text{sin}}^{\beta=2}(x_1,x_2)&=\frac{\sin(x_1-x_2)}{\pi(x_1-x_2)}=
\frac{1}{\pi} \int_0^1 
dt\ \cos[(x_1-x_2)t]\ ,
\label{SinKb2}\\ 
&&\nn\\
&& K_{\text{sin},\mathbb C}^{\beta=2}(z_1,z_2)&=\frac{1}{\sig\pi^{3/2}}\ 
\e^{-\frac{y_1^2+y_2^2}{2\sig^2}}\int_0^1 dt\ \e^{-\sig^2
  t^2}\cos[(z_1-z_2)t]\ . 
\ \label{SinKb2C}
\end{align}
The corresponding spectral density of complex eigenvalues 
was first derived in \cite{FKS} using supersymmetry, 
and the kernel with all correlations functions in \cite{FKS98} using
OP, see eq.\ (\ref{Kb2C}). 
In the Hermitian limit $\sigma\rightarrow 0$, we have 
$K_{\text{sin},\mathbb  C}^{\beta=2}(z_1,z_2) \rightarrow 
\sqrt{\delta^{1}(y_1)\delta^{1}(y_2)}
K_{\text{sin}}^{\beta=2}(x_1,x_2)$.

In \cite{FKS98} it was shown using
supersymmetric techniques that the same result holds for the
microscopic density of random
matrices with i.i.d.\ matrix elements for $\beta=1,2$. Further
arguments in favour 
of universality were added in \cite{Ake02} for the kernel using large-$N$
factorisation and asymptotic OP. The universal parameter is the mean
macroscopic spectral density $\rho(x_0)$.
\begin{align}
\underline{\beta=4:}
&& G_{\text{sin}}^{\beta=4}(x_1,x_2)&=
-\, \frac{\sin[2(x_1-x_2)]}{2\pi(x_1-x_2)},
\ K_{\text{sin}}^{\beta=4}(x_1,x_2)=
\frac{\partial}{\partial x_1}
G_{\text{sin}}^{\beta=4}(x_1,x_2),
\nn\\
&& W_{\text{sin}}^{\beta=4}(x_1,x_2)&=\int_0^{x_1-x_2} dt\
G_{\text{sin}}^{\beta=4}(t,0)
=\frac{1}{2\pi} \int_0^1 \frac{dt}{t} \sin[2(x_1-x_2)t]\ ,
\label{SinKb4}\\
&&\nn\\
&& G_{\text{sin},\mathbb C}^{\beta=4}(z_1,z_2) &=  
\frac{i2\sqrt{2}\,y_2}{\pi^{3/2}\sigma^3}
\e^{-\frac{2y_2^2}{\sigma^2}}
\int_0^1 \frac{dt}{t}\ \e^{-2\sig^2 t^2}\sin[2(z_1-z_2^*)t]\ .
\label{SinKb4C}
\end{align}
The corresponding spectral density of complex eigenvalues was derived
in \cite{KE} using supersymmetry, and the kernel with all correlations
functions was derived in \cite{EK}
using skew-OP leading to eq.\ (\ref{Kb4C}).
\begin{align}
\underline{\beta=1:}
&& G_{\text{sin}}^{\beta=1}(x_1,x_2)&=
-K_{\text{sin}}^{\beta=2}(x_1,x_2)\ ,\nn\\ 
&&K_{\text{sin}}^{\beta=1}(x_1,x_2)&=\frac{\partial}{\partial x_1}
G_{\text{sin}}^{\beta=1}(x_1,x_2)
=\frac{1}{\pi} \int_0^1 dt \, t\ \sin[(x_2-x_1)t]
\ ,
\nn\\
&& W_{\text{sin}}^{\beta=1}(x_1,x_2)&=\int_0^{x_1-x_2} dt\
G_{\text{sin}}^{\beta=1}(t,0) + \tfrac12 \sign(x_1-x_2)\ ,
\label{SinKb1}\\
&&\nn\\
&& G_{\text{sin},{\mathbb C},\text{real}}^{\beta=1}(x_1,x_2)&=-\, \frac{1}{\pi}
\int_0^1 dt\ \e^{-\sig^2 t^2}\cos[(x_1-x_2)t]\ ,\nn\\
&& G_{\text{sin},{\mathbb C},\text{com}}^{\beta=1}(z_1,z_2)&=-\, 
2i \, \sign(y_2)
\erfc\left(\frac{|y_1|}{\sig}\right) 
K_{\text{sin},\mathbb  C}^{\beta=1}(z_1,z_2^*)\ ,\nn\\
&& K_{\text{sin},\mathbb C}^{\beta=1}(z_1,z_2)&=
\frac{1}{\pi} \int_0^1 dt \, t\ \e^{-\sig^2 t^2}\sin[(z_2-z_1)t]\ .
\label{SinKb1C}
\end{align}
The kernel elements
$G_{\text{sin},{\mathbb C}}^{\beta=1}(z_1,z_2)$ and
$K_{\text{sin},\mathbb  C}^{\beta=1}(z_1,z_2)$ 
were derived in \cite{Forrester07} using skew-OP, c.f. eq.\ (\ref{Kb1C}).
The same resulting spectral densities of complex and real eigenvalues 
were derived previously in \cite{Efetov97} using a sigma-model
calculation, which again indicates universality. It can easily be verified here
that the Hermitian limit $\sig\to 0$ of $G_{\text{sin},{\mathbb
    C},\text{real}}^{\beta=1}(x_1,x_2)$   
is indeed $G_{\text{sin}}^{\beta=1}(x_1,x_2)$, and that
$G_{\text{sin},{\mathbb C},\text{com}}^{\beta=1}(z_1,z_2)$  
vanishes in this limit.

\subsection{Wishart-Laguerre ensembles with eigenvalues 
on ${\mathbb R}$ and ${\mathbb C}$}
\label{WL}

In order to be able to access the Bessel kernels for real and
complex eigenvalues as well, we briefly introduce the Wishart-Laguerre
(or chiral) ensembles (L$\beta$E) and their non-Hermitian counterparts 
(${\mathbb C}$L$\beta$E).
We begin with the former which are defined as 
\be
\mcZ_{N}^{\text{L$\beta$E}}= \int dW \exp[-\beta\Tr WW^\dag/2]
=c_{N,\beta,\nu}
\prod_{j=1}^N  \int_{{\mathbb R}_+}
dx_j\,w_{\beta}^\nu(x_j)\ |\Delta_N(\{x\})|^\beta\ .  
\label{LbE}
\ee
The elements of the rectangular $N\times (N+\nu)$ matrix $W$ are
again real, complex, or quaternion real for $\beta=1,2,4$, without
further symmetry constraints. The integration denoted by $dW$ runs
over all the independent matrix elements. Because we want to access the
so-called hard edge of the spectrum 
we will only consider fixed $\nu=O(1)$ in the following. The
distribution of the positive definite eigenvalues $x_j$ of $WW^\dag$ in the
Wishart picture (or
equivalently the distribution of the singular values of $W$ in the
Dirac picture used in QCD)  
is of the same form as
eq.\ (\ref{GbE}), but with different weight functions 
\be
w_{\beta}^\nu(x)=x^{\frac12\beta(\nu+1)-1} \exp[-\beta x/2]\ ,
\label{wLaguerre}
\ee
that now depend on $\beta$ in a non-trivial way.
Consequently the $k$-point correlation functions take the same form as
in eq.\ (\ref{Rkreal}), with the corresponding kernels. 

In analogy to the Ginibre ensembles we define a parameter-dependent
family of non-Hermitian Wishart-Laguerre (also called complex chiral) ensembles
as the following two-matrix model 
\be
\mcZ_N^{\text{${\mathbb C}$L$\beta$E}}(\tau)=\int dW dV 
\exp\left[- \frac{1}{1-\tau}\Tr\Big( 
WW^\dag+ V^\dag V-\tau(WV+V^\dag W^\dag)\Big)\right], 
\label{CLbE}
\ee
with $W$ and $V^\dag$ being two rectangular $N\times (N+\nu)$ matrices.
Here we follow the notation of \cite{ABe}. This two-matrix model was 
first introduced and solved for $\beta=1,2,4$ in \cite{APSo,AOSV,A05}
respectively. 
For $\tau=0$ the jpdf of all the matrix elements again factorises, and
in the opposite 
limit we have
$\mcZ_N^{\text{${\mathbb C}$L$\beta$E}}(\tau) \rightarrow
\mcZ_N^{\text{L$\beta$E}}$ 
as $\tau \to 1$.
Here we are seeking the complex (and real) eigenvalues of the product
matrix $WV$ ($W$ and $V$ are the off-diagonal blocks of the Dirac
matrix that we diagonalise). 
Its jpdf takes the same form as in eqs.\ (\ref{GinUE})
and (\ref{GinSOE}), but with different weight functions that are no
longer Gaussian 
\begin{align}
w_{\beta=2}^{\nu,\,\mathbb  C}(z) &= |z|^{\nu}
\exp\left[\frac{\tau(z+z^*)}{1-\tau^2}\right]
K_\nu\left(\frac{2|z|}{1-\tau^2}\right),\nn\\
  w_{\beta=4}^{\nu,\,\mathbb
    C}(z) &= \sqrt{w_{\beta=2}^{2\nu,\,\mathbb  C}(z)}\ .
\label{wLaguerreC}
\end{align}
For $\beta=4$ the anti-symmetric weight function is defined as in
eq.\ (\ref{Fb4}).  
For $\beta=1$ we explicitly specify two functions in 
the anti-symmetric weight function
\begin{align}
{\cal F}_{\beta=1}^{\nu,\mathbb C} (z_{1},z_{2})
&={i}g_\nu(z_{1},z_{2}) \sign(y_1) \,
\delta^2(z_1-z_2^*)
+  \tfrac12  h_\nu(x_1)h_\nu(x_2)\delta(y_{1})\delta(y_{2})
\sign(x_{2}-x_{1}) \, ,
\label{Fb1ch}
\nn\\
h_\nu(x) &=
2|x|^{\frac{\nu}{2}}\exp\left[\frac{\tau x}{1-\tau^2}\right]
K_{\frac{\nu}{2}}\left(\frac{|x|}{1-\tau^2}\right) 
\ ,\\
g_\nu(z_1,z_2)&=
2|z_1z_2|^{\frac{\nu}{2}}\,
\exp\left[ \frac{\tau(z_1+z_2)}{1-\tau^2} \right] \nn \\
 & \qquad \times \int_0^\infty \frac{dt}{t}\,\exp\left[ - \,
  \frac{(z_1^2+z_2^2)t}{(1-\tau^2)^2}-\frac{1}{4t} \right] 
K_{\frac{\nu}{2}}\left(\frac{2 z_1 z_2 t}{(1-\tau^2)^2} \right)
\erfc\left( \frac{|z_2-z_1|\sqrt{t}}{1-\tau^2} \right),
\nn
\end{align}
which are related by $g_\nu(z,z^*) \rightarrow h_\nu(x)^2$ as $y\to0$.
We give the corresponding kernels in Appendix \ref{SOP}.

In the large-$N$ limit (with $\nu=O(1)$ fixed),
for all three $\beta$ the real positive Wishart
eigenvalues of $WW^{\dagger}$ are concentrated on the interval $(0,4N]$, with a 
density $\rho(x)=(2\pi N)^{-1}\sqrt{(4N-x)/x}$. 
This is a special case of the
Marchenko-Pastur density. After mapping to Dirac
eigenvalues $\la = \sqrt{x}$,
this becomes the same semi-circle distribution as for WD,
but with eigenvalues coming in
$\pm\la$ pairs, together with $\nu$ exactly zero eigenvalues making the origin
special.
The density of the complex Wishart eigenvalues has a
singularity at the origin; however, after mapping to the Dirac picture,
we obtain a macroscopic density function that is flat on an ellipse,
just as in the Ginibre case.  

We now give a list of all the known Bessel kernels. For real eigenvalues
we follow \cite{DGKV,NF} where a most comprehensive list and references
can be found. In some cases the parallel between kernels of real and
complex eigenvalues is more transparent after using some identities
for Bessel functions.

\subsection{Limiting Bessel kernels on ${\mathbb R}$ and ${\mathbb
    C}$}\label{secLimitingBessel} 

The hard-edge limit is defined by zooming into the origin
(see Appendix \ref{evalscalings}), where for the
complex eigenvalues we have to keep $\sig=\sqrt{N(1-\tau)}$
fixed as in the weakly non-Hermitian bulk limit
eq.\ (\ref{sig-sine}). The corresponding limiting kernels are given as follows:
\begin{align}
\underline{\beta=2:}
&& K_{\text{Bes}}^{\beta=2}(x_1,x_2)&=\frac{J_\nu(\sqrt{x_1})\sqrt{x_2}\,
J_{\nu-1}(\sqrt{x_2})-(x_1 \leftrightarrow x_2)}
{2(x_1-x_2)}
=\tfrac12\int_0^1 dt \, t \,
J_\nu(\sqrt{x_1}\,t)J_\nu(\sqrt{x_2}\,t)\ , \ \ \ \ \nn \\
\label{BesKb2}\\
&& K_{\text{Bes},\mathbb C}^{\beta=2}(z_1,z_2)&=
\frac{1}{8\pi\sig^2} 
K_\nu\left(\frac{|z_1|}{4\sig^2}\right)^{\frac12}
K_\nu\left(\frac{|z_2|}{4\sig^2}\right)^{\frac12}
\e^{\frac{x_1+x_2}{8\sigma^2}}
\int_0^1 dt\,t \, \e^{-2\sig^2 t^2}J_\nu(t\sqrt{z_1})
J_\nu(t\sqrt{z_2})\ .\nn\\
\label{BesKb2C} 
\end{align}
It can be shown that
$K_{\text{Bes},\mathbb C}^{\beta=2}(z_1,z_2) \rightarrow
\sqrt{\delta^{1}(y_1)\delta^{1}(y_2)} \Theta(x_1)\Theta(x_2)
K_{\text{Bes}}^{\beta=2}(x_1,x_2)$ as $\sigma\rightarrow 0$.
The kernel of complex eigenvalues was derived in
\cite{AOSV}. The same density following from this kernel was
obtained from a different Gaussian
non-Hermitian one-matrix model \cite{KS} using replicas, 
and is in that sense universal.
\begin{align}
\underline{\beta=4:}
&& K_{\text{Bes}}^{\beta=4}(x_1,x_2)&=
2\int_0^1dt\, t \int_0^1 ds \, s^3 
\Big(J_{2\nu+1}(2\sqrt{x_1}\,st)J_{2\nu+1}(2\sqrt{x_2}\,s)
-(x_1 \leftrightarrow x_2)
\Big),
\nn\\
&& G_{\text{Bes}}^{\beta=4}(x_1,x_2)&=-
2\sqrt{x_1}\int_0^1dt \int_0^1 ds \, s^2 
\Big(J_{2\nu}(2\sqrt{x_1}\,st)J_{2\nu+1}(2\sqrt{x_2}\,s)
-tJ_{2\nu}(2\sqrt{x_1}\,s)J_{2\nu+1}(2\sqrt{x_2}\,st)\Big),
\nn\\
&& W_{\text{Bes}}^{\beta=4}(x_1,x_2)&=
2\sqrt{x_1x_2}\int_0^1dt\int_0^1
ds \, s \, \Big(J_{2\nu}(2\sqrt{x_1}\,st)J_{2\nu}(2\sqrt{x_2}\,s)
-(x_1\leftrightarrow x_2)
\Big),
\label{BesKb4}\\
&&\nn\\
&& K_{\text{Bes},\mathbb C}^{\beta=4}(z_1,z_2) &= 
\frac{1}{\sig^4} \int_0^1 dt\int_0^1 ds \, s  \,
\e^{-2\sig^2 s^2(1+t^2)}\Big(J_{2\nu}(2\sqrt{z_1}\,st)J_{2\nu}(2\sqrt{z_2}\,s)
-(z_1\leftrightarrow z_2)
\Big)\ .
\label{BesKb4C}
\end{align}
The complex kernel was first derived in \cite{A05}, whereas
the matching in the Hermitian limit -- which can best be seen when
comparing the kernels $K_{\text{Bes},\mathbb C}^{\beta=4}$ 
in eq. (\ref{BesKb4C}) with $W_{\text{Bes}}^{\beta=4}$ in
eq. (\ref{BesKb4}) -- is
discussed in detail in \cite{ABa}. 
One has to take the weight function into account when taking
the Hermitian limit.

\begin{align}
\underline{\beta=1:}
&& K_{\text{Bes}}^{\beta=1}(x_1,x_2)&=
\frac{-1}{8\sqrt{x_1x_2}}
 \int_0^1 ds\, s^2
\left\{\sqrt{x_1}\,J_{\nu+1}(s\sqrt{x_1})J_{\nu}(s\sqrt{x_2}) - (x_1
\leftrightarrow x_2)\right\}
=-
\frac{\partial}{\partial x_2}G_{\text{Bes}}^{\beta=1}(x_1,x_2),
\nn\\
&& G_{\text{Bes}}^{\beta=1}(x_1,x_2)&=-\,
\tfrac12\int_0^1 dt \, t \, J_{\nu-1}(\sqrt{x_1}\,t)J_{\nu-1}(\sqrt{x_2}\,t)
-\frac{1}{4\sqrt{x_1}}
J_\nu(\sqrt{x_1})\int_{\sqrt{x_2}}^\infty
ds J_{\nu-2}(s)\ ,\nn\\
&& W_{\text{Bes}}^{\beta=1}(x_1,x_2)&=
-\int_{x_1}^{x_2}ds\, G_{\text{Bes}}^{\beta=1}(s,x_2)
-\tfrac12\sign(x_1-x_2)\ ,
\label{BesKb1}
\end{align}
\begin{align}
K_{\text{Bes},\mathbb C}^{\beta=1}(z_1,z_2)&
 = \frac{1}{256 \pi\sigma^2} \int_0^1 \,ds\, s^2\,\e^{-2\sigma^2 s^2}\,
\left\{\sqrt{z_1}\,J_{\nu+1}(s\sqrt{z_1})J_{\nu}(s\sqrt{z_2}) - (z_1
\leftrightarrow z_2) \right\}\ ,\nn\\
G_{\text{Bes},\mathbb C,\text{com}}^{\beta=1}(z_1,z_2) &= 
 -2i \, \sign(y_2)\,\e^{\frac{x_2}{4 \sigma^2}}\! 
  \int_0^\infty\!
  \frac{dt}{t}\,\e^{-\frac{t}{64\sigma^4}
  (z_2^2+z_2^{*\,2})-\frac{1}{4t}}
  K_{\frac{\nu}{2}}\Big(\frac{t}{32\sigma^4}|z_2|^2\Big)\nn\\
 & \qquad \times \erfc\Big(\frac{\sqrt{t}\,|y_2|}{4\sigma^2}\Big)
K_{\text{Bes},\mathbb C}^{\beta=1}(z_1,z_2^*)\ ,
\nn\\
G_{\text{Bes},\mathbb C,\text{real}}^{\beta=1}(x_1,x_2) &= -
  \frac{2\,\e^{\frac{x_2}{8\sigma^2}}
  K_{\frac{\nu}{2}}\left( \frac{|x_2|}{8\sigma^2}
  \right)}{[\sign(x_2)]^{\frac{\nu}{2}}}  
  \Bigg\{\!\! \left( (-i)^{\nu} \int_{-\infty}^0 dy +
  \frac{2}{[\sign(x_2)]^{\frac{\nu}{2}}} \int_{0}^{x_2} dy \right) 
  K_{\text{Bes},\mathbb C}^{\beta=1}(x_1,y) 
 \nn \\
 & \ \   \times 2\,\e^{\frac{y}{8\sigma^2}}K_{\frac{\nu}{2}}\left(
 \frac{|y|}{8\sigma^2} \right) 
- \frac{1}{32\sqrt{\pi}} 
  \Bigg[ - \, \frac{1}{\sigma}\,\e^{-\sigma^2}\,J_{\nu}(\sqrt{x_1})
  + \frac{2 \sigma^{\nu}}{\Gamma\left(\frac{\nu+1}{2}\right)}\,
  \int_0^1 ds\,\e^{-\sigma^2 s^2} s^{\nu+2} 
  \nn
  \\
  & \ \   \times\! \left(
  \frac{\sqrt{x_1}}{2}\,E_{\frac{1-\nu}{2}}(\sigma^2 s^2)\,
  J_{\nu+1}(s \sqrt{x_1}) - \sigma^2s\left(
  E_{\frac{-1-\nu}{2}}(\sigma^2 s^2) - E_{\frac{1-\nu}{2}}(\sigma^2 s^2) \right)
  J_{\nu}(s \sqrt{x_1}) \right)\!\! \Bigg]\!
  \Bigg\}\,.
\nn\\
\label{BesKb1C}
\end{align}
In the final equation above, $\displaystyle E_n(x) = \int_{1}^{\infty}
dt \, \frac{\e^{-xt}}{t^n}$ is the exponential integral. 
The kernels 
$K_{\text{Bes},\mathbb C}^{\beta=1}(z_1,z_2)$ and 
$G_{\text{Bes},\mathbb C,\text{real}}^{\beta=1}(x_1,x_2)$
were derived in \cite{APSo}, whereas the non-commutativity issues of 
$G_{\text{Bes},\mathbb C,\text{real}}^{\beta=1}(x_1,x_2)$ in the weak limit can
be found in \cite{AKPW,Michael}.
There is numerical evidence from studying some examples of
non-Gaussian RMT that the density of the real eigenvalues resulting 
from $G_{\text{Bes},\mathbb C,\text{real}}^{\beta=1}(x_1,x_2)$ and the
corresponding distribution of the smallest eigenvalues may be universal
\cite{Michael}. 

The Hermitian limit is much more involved here (compare however 
$K_{\text{Bes},\mathbb C}^{\beta=1}(z_1,z_2)$ and
$K_{\text{Bes}}^{\beta=1}(z_1,z_2)$); in particular, as 
$N\to\infty$ and $\sigma\to 0$, we find that
$G_{\text{Bes},\mathbb C,\text{com}}^{\beta=1}(z_1,z_2) \rightarrow 0$
and $G_{\text{Bes},\mathbb C,\text{real}}^{\beta=1}(x_1,x_2) \rightarrow
G_{\text{Bes}}^{\beta=1}(x_1,x_2)$, and we refer to
\cite{AKPW,Michael} for more details. 

\sect{Discussion and open problems}\label{conc}

In this short article we have collected together all the known kernels for RMT
with real eigenvalues along with all the known and new kernels for RMT
with complex eigenvalues at weak non-Hermiticity. This comprises the
Airy, sine and Bessel kernels of the three Wigner-Dyson and the three
Wishart-Laguerre ensembles, as well as their non-Hermitian
counterparts. In order to highlight the nature of this deformation 
we have used real integral representations for the kernels of real
eigenvalues, rather than the asymptotic forms resulting from the
Christoffel-Darboux identity for $\beta=2$ or from the rewriting \`a la
Tracy-Widom for $\beta=1,4$. The extra exponential factor (and shift
for the Airy case) in the integral representation of the kernels on
$\mathbb C$ is a very smooth deformation. This makes it very plausible that
the universality which is very well studied for real eigenvalues 
extends to the weakly non-Hermitian limit for all
ensembles, beyond what is already known for $\beta=2$.
The universality of the factor in front of the integral which
contains special functions such as the complementary error function or
modified Bessel function will be more 
difficult to establish. However, the presence of these factors is crucial 
when taking the Hermitian limit, in projecting the imaginary parts of the
eigenvalues to zero.

Whilst we have already mentioned what is known about universality in
the weak limit so far, let us give some more open problems. To date, a
mathematically rigorous derivation of most of the limiting kernels on $\mathbb
C$ is lacking, apart from the complex Airy kernel for $\beta=2$
\cite{Bender}. There is no doubt that the kernels we have listed
and which have been derived using different techniques such as
asymptotic OP, supersymmetry or replicas are correct. This is 
based not only on numerical evidence but also, and more importantly,
on a comparison with complex 
eigenvalue spectra in physics, see  e.g.\ \cite{Jac} and references
therein, where the complex Bessel kernels for $\beta=2$ and $4$ were
successfully compared 
with complex spectra from QCD and QCD-like theories. Because the latter
are field theories and not Gaussian RMT this gives a further
indication that universality
holds in this regime. Preliminary numerical
investigations with non-Gaussian, non-Hermitian RMT 
appear  to confirm this \cite{Michael} 
for $\beta=1$. 

A much more challenging problem will be to show the universality of
these kernels on $\mathbb C$, either by going to non-Gaussian potentials of
polynomial or harmonic form, or by considering non-Hermitian Wigner 
matrix ensembles, with elements being
independent random variables. 

A further reason why we believe that this universality
question is important is that some of the kernels on $\mathbb C$
reappear in the same integral form (with real arguments) 
when looking at symmetry
transitions between two different {\it Hermitian} RMTs, say from one GUE
to another GUE, in a corresponding ``weak" limit. 
Their eigenvalue
correlations are  
also called parametric. For $\beta=2$ this fact can 
be observed for the Bessel, sine \cite{ADOS,peter} and Airy 
\cite{Macedo,peter} kernels.

\vspace{0.5cm}
{\sc Acknowledgements}:~
The organisers and participants of the workshop
``Random Matrix Theory and its Applications'' at MSRI Berkeley,
13th-17th September 2010, 
are thanked for many inspiring talks and discussions.

\appendix

\sect{Finite-$N$ (skew-) orthogonal 
polynomials 
and kernels on ${\mathbb C}$}\label{SOP}

In this appendix we specify the orthogonal polynomials (OP) and skew-OP as
well as their (skew-) symmetric scalar products. These may be used to
construct the 
kernels -- which we will list for all the above matrix ensembles with
complex eigenvalues -- 
in terms of which all $k$-point correlation functions can be
expressed, see eqs.\ (\ref{Rkreal}) and (\ref{Rkcomp}). We also highlight
the relations between expectation values of characteristic polynomials
on the one hand and (skew-) OP and their kernels on the other,
valid on both ${\mathbb R}$ and ${\mathbb C}$.

Starting with $\beta=2$ we define the monic OP on ${\mathbb R}$ (and
${\mathbb C}$) by
\be
\int_{\mathbb R (\mathbb C)} d^{(2)}z\ w_{\beta=2}^{(\nu,{\mathbb C})}(z)
P_k(z)P_l(z)^{(*)}=h_{k(\nu,{\mathbb C})}^{\beta=2}\delta_{kl} \ ,
\label{OPdef}
\ee
with squared norms $h_{k(\nu,{\mathbb C})}^{\beta=2}$. 
Because in our examples all
moments exist these OP can be constructed via
the Gram-Schmidt procedure. Alternatively, they can be written as 
\be
P_k(z)=\Big\langle \det[z-H]\Big\rangle_k=\frac{1}{\mcZ_{k}^{\text{GUE}}}
\int dH \det[z-H]\exp[-\beta\Tr H^2/4]\ ,
\label{OPvev}
\ee
and similarly for the GinUE and ($\mathbb C$)LUE, replacing the
$k\times k$ matrix $H$ with $J$,
or the Wishart matrices $WW^\dag$ and $WV$ 
respectively. In fact, this relation holds for
general weight functions. For the Gaussian ensembles we obtain
Hermite, and for the WL ensembles Laguerre polynomials
on ${\mathbb R}$ and on ${\mathbb C}$. The corresponding kernels are 
then obtained by summing over the {\it normalised} OP (multiplied by the
weights). Most conveniently, a second relation to 
characteristic polynomials exists \cite{AV03},
\be
K_{N,\mathbb C}^{\beta=2}(u,v)= w_{\beta=2}^{{\mathbb C}}(u)^{\frac12}
 w_{\beta=2}^{{\mathbb C}}(v)^{\frac12}
\frac{1}{h_{N-1,{\mathbb C}}^{\beta=2}}\Big\langle \det[u-J]\det[v-J^\dag]
\Big\rangle_{N-1}\ ,
\label{Kervev}
\ee
which we state here for the GinUE. Correspondingly it holds for the GUE and 
$\beta=2$ WL ensembles, and, indeed, for arbitrary weights. We can now
give the two 
$\beta=2$ kernels in the complex plane, following \cite{FKS98} and
\cite{AOSV} respectively:
\begin{align}
\underline{\beta=2:}
&& K_{N,\mathbb C}^{\beta=2}(u,v)&= w_{\beta=2}^{{\mathbb C}}(u)^{\frac12}
 w_{\beta=2}^{{\mathbb C}}(v)^{\frac12}\frac{1}{\pi\sqrt{1-\tau^2}}
\sum_{j=0}^{N-1}\frac{\tau^j}{2^j j!}
H_{j}\Big(\frac{u}{\sqrt{2\tau}}\Big)H_{j}\Big(\frac{v}{\sqrt{2\tau}}\Big), 
\label{Kb2C}\\
&& K_{N,\nu,\mathbb C}^{\beta=2}(u,v)&= w_{\beta=2}^{\nu,{\mathbb
    C}}(u)^{\frac12} 
 w_{\beta=2}^{\nu,{\mathbb C}}(v)^{\frac12}\frac{2}{\pi(1-\tau^2)}
\sum_{j=0}^{N-1}
\frac{\tau^{2j}j!}{(j+\nu)!}L_{j}^\nu\Big(\frac{u}{\tau}\Big)
L_{j}^\nu\Big(\frac{v}{\tau}\Big)\ .
\label{Kb2Cch}
\end{align}

For the skew-OP related to $\beta=1,4$, we have to distinguish between
the skew products for complex and real eigenvalues. Because the latter
are very well known (see e.g.\ \cite{Mehta}) 
we will focus on the former, which can be written in a unified way
\cite{AKP}
\be
\int_{{\mathbb C}^2} \dsq z_1 \dsq z_2\ 
{\cal F}^{(\nu){\mathbb C}}_{\beta=1,4}(z_1,z_2)\det\left[
\begin{array}{cc}
Q_{2k}^{\beta=1,4}(z_1) & Q_{2l+1}^{\beta=1,4}(z_1)\\
Q_{2k}^{\beta=1,4}(z_2) & Q_{2l+1}^{\beta=1,4}(z_2)\\
\end{array}
\right]=h_{k,(\nu){\mathbb C}}^{\beta=1,4}\delta_{kl} \ ,
\label{skewdef}
\ee
for skew-OP of even-odd degree, and which is vanishing 
for even-even and odd-odd degree.
Here and in the following we again choose $N$ even.
Once more, the skew-OP satisfying this can be written as follows for
$\beta=4$ \cite{EK} and $\beta=1$ \cite{AKP}
\be
Q_{2k}^{\beta=1,4}(z)=\Big\langle \det[z-J]\Big\rangle_{2k}\ ,\ \ 
Q_{2k+1}^{\beta=1,4}(z)=\Big\langle \det[z-J](z+c+\Tr J)\Big\rangle_{2k}.
\label{sOPdef}
\ee
Note that, for $\beta=4$, the matrix $J$ here should be taken as
the complex-valued matrix of size $2k \times 2k$.
The odd skew-OP in eq.\ (\ref{sOPdef}) are defined only up to a
constant $c$ times the 
even skew-OP. The same
relation holds for real eigenvalues and arbitrary weights, see
\cite{AKP} for references.
Moreover, the anti-symmetric kernel matrix element $K_{N,\mathbb C}^{\beta=1,4}$
(sometimes called the pre-kernel) enjoys a similar
relation to that in eq.\ (\ref{Kervev}), as was observed for $\beta=4$
\cite{ABa} and $\beta=1$ \cite{APSo}
\be
K_{N,\mathbb C}^{\beta=1,4}(u,v)= 
(u-v)
\frac{1}{h_{\frac{N}{2}-1,{\mathbb C}}^{\beta=1,4}}\Big\langle
\det[u-J]\det[v-J^\dag] 
\Big\rangle_{N-2} \, .
\label{qKervev}
\ee
We list the corresponding kernel matrix elements following \cite{EK}
and \cite{A05} respectively
\begin{align}
\underline{\beta=4:}
&& K_{N,\mathbb C}^{\beta=4}(u,v)&=
\frac{1}{\pi(1-\tau)\sqrt{1-\tau^2}} 
\sum_{k=0}^{N/2-1}\sum_{l=0}^k\frac{1}{(2k+1)!!(2l)!!} \,
\left(\frac{\tau}{2}\right)^{k+l+\frac12} \nn \\ 
&& & \qquad \times \left(H_{2k+1}\left(\frac{u}{\sqrt{2\tau}}\right)
H_{2l}\left(\frac{v}{\sqrt{2\tau}}\right) 
-(u\leftrightarrow v)\right)\ ,
\label{Kb4C}\\
&& K_{N,\nu\mathbb C}^{\beta=4}(u,v) &= - \, \frac{2}{\pi(1-\tau^2)^2}
\sum_{k=0}^{N/2-1}\sum_{j=0}^k
\frac{2^{2k-2j}k!(k+\nu)!(2j)!}
{(2k+2\nu+1)!j!(j+\nu)!} \tau^{2k+2j+1} \nn \\
&& & \qquad \times \left( 
L_{2k+1}^{2\nu}\left(\frac{u}{\tau}\right)
L_{2j}^{2\nu}\left(\frac{v}{\tau}\right)
-(u\leftrightarrow v)
\right),
\label{Kb4Cch}
\end{align}
(recalling that $N$ here is the size of the complex-valued matrix that
is equivalent to the original quaternion real matrix)
and \cite{Forrester07} and \cite{APSo}
\begin{align}
\underline{\beta=1:}
\!\!\!\!\!\!&& K_{N,\mathbb C}^{\beta=1}(u,v)&=
\ \frac{1}{2\sqrt{2\pi}(1+\tau)} \sum_{l=0}^{N-2}\frac{1}{l!}
\left( \frac{\tau}{2}\right)^{l+\frac{1}{2}}\left( 
H_{l+1}\left( \frac{u}{\sqrt{2\tau}}\right)
H_l\left( \frac{v}{\sqrt{2\tau}}\right)
-(u\leftrightarrow v)\right),
\label{Kb1C}\\
&& K_{N,\nu,\mathbb C}^{\beta=1}(u,v)&=
- \frac{1}{8\pi(1-\tau^2)}
\sum_{l=0}^{N-2}
\frac{(l+1)!}{(l+\nu)!} \,
\tau^{2l+1}
\left(
L_{l+1}^\nu\Big( \frac{u}{\tau}\Big)
L_{l}^\nu\Big( \frac{v}{\tau}\Big)
-(u\leftrightarrow v)
\right).
\label{Kb1Cch}
\end{align}
The other elements of the matrix-valued kernel follow by integration.

\section{Large-$N$ limits at weak non-Hermiticity}\label{evalscalings}

In this appendix we will specify the different large-$N$ limits that
lead to the limiting kernels listed in Section
\ref{results}.
Let us emphasise that these are not all of the possible
large-$N$ limits of the above finite-$N$ kernels that one can take.
We will give only those limits where $(1-\tau)N^\delta=\sigma$ is kept
fixed for some $\delta>0$,
limits where the degree of non-Hermiticity is weak.
The reason is that it is only these particular limiting kernels that
relate closely to the known universal kernels on $\mathbb R$.
However, many of the results at strong non-Hermiticity (i.e.\ where $\tau$ is
$N$-independent) can be recovered from the weak limit by taking
$\sigma\to\infty$ and rescaling the complex eigenvalues accordingly.\\

\noindent
{\bf Soft edge limit:}
We consider fluctuations around the right end-point of the long half-axis of
the supporting ellipse, 
to obtain from eq.\ (\ref{Kb2C}) \cite{Bender}
\be
z = (1+\tau)\sqrt{N}+\frac{X}{N^{1/6}}\ +\ i\ \frac{Y}{N^{1/6}}\ ,\ \ \ 
\sig= N^{1/6}\sqrt{1-\tau}\ ,
\label{limAi}
\ee
\be
K_{\text{Ai},\mathbb  C}^{\beta=2}(X_1+iY_1,X_2+iY_2) \equiv
\lim_{\substack{N\to\infty \\ \tau\to1}} 
\frac{1}{N^{1/3}}K_{N,\mathbb C}^{\beta=2}(z_1,z_2)\ .
\label{limAiK}
\ee
The same limit applies to the WL kernel
on $\mathbb C$ given in eq.\ (\ref{Kb2Cch}) \cite{ABe}. By symmetry we expect
the same limiting behaviour around the left end-point 
$-(1+\tau)\sqrt{N}$
as well as for $\nu=O(N)$. 
The limiting kernels for $\beta=1,4$ are defined
in the same way.
Note that the eigenvalues in the bulk of
the spectrum are at strong non-Hermiticity in this limit, since
$\sig_{\text{sine}} = N^{1/3}\sig_{\text{Airy}} \to \infty$ as $N\to\infty$. 
In fact, in order to reach weak non-Hermiticity in the bulk we need to consider
the following scaling limit.\\

\noindent
{\bf Bulk limit:}
Without loss of generality we consider fluctuations around the origin,
being representative of the Gaussian ensembles eq.\ (\ref{GinbE}).
On rescaling, we obtain from eq.\ (\ref{Kb2C}) \cite{FKS}
\be
z=\frac{X}{N^{1/2}}\ +\ i\ \frac{Y}{N^{1/2}}\ ,\ \ \ 
\sig = N^{1/2}\sqrt{1-\tau}\ ,
\label{limsin}
\ee
\be
K_{\text{sin},\mathbb  C}^{\beta=2}(X_1+iY_1,X_2+iY_2) \equiv
\lim_{\substack{N\to\infty \\ \tau\to1}} 
\frac{1}{N}K_{N,\mathbb C}^{\beta=2}(z_1,z_2)\ ,
\label{limsinK}
\ee
and likewise for $\beta=1,4$. The macroscopic spectral
density collapses onto the real axis, and becomes the semi-circle for
our Gaussian ensembles. The functions
$R_{k,\mathbb C}$ given by the determinant or Pfaffian of the rescaled
kernels describe the microscopic correlations in the complex plane.

If we magnify around any other point $|x_0|<2\sqrt{N}$ inside the bulk,
then we rescale
the fluctuations $z-x_0$ as in eq.\ (\ref{limsin}). The correlations are
then universal when measured in units of the local mean density
$\pi\rho_{\text{sc}}(x_0)$. 
\\

\noindent
{\bf Hard edge limit:} Whilst we expect that in the bulk of the
spectrum the WL and Gaussian ensembles (eqs.\ (\ref{CLbE}) and
(\ref{GinbE}) respectively) show the same behaviour, the 
origin is singled out in the latter case. The rescaling here is given by \cite{AOSV}
\be 
z = \frac{X}{4N} + i \, \frac{Y}{4N},\ \ \ 
\sig = N^{1/2}\sqrt{1-\tau}\ ,
\ee
\be
K_{\text{Bes},\mathbb  C}^{\beta=2}(X_1+iY_1,X_2+iY_2) \equiv
\lim_{\substack{N\to\infty \\ \tau\to1}} 
\frac{1}{(4N)^2}K_{N,\nu,\mathbb C}^{\beta=2}(z_1,z_2)\ .
\label{limBesK}
\ee
with the Laguerre polynomials in
eq.\ (\ref{Kb2Cch}) displaying a
Bessel function asymptotic.
\\

In all three scaling limits the asymptotic kernels are obtained by
replacing the sums with integrals (the Christoffel-Darboux identity does
not hold for OP in the complex plane), and the Hermite and
Laguerre polynomials by their corresponding Plancherel-Rotach
asymptotics (proven for real arguments) in the corresponding region.

An additional problem arises from the integrations with the
anti-symmetric weight function ${\cal F}$ used to obtain the limiting
kernel elements $G$ and 
$W$. For $\beta=1$ these integrals are not absolutely
convergent, and hence the limit $N\to\infty$ and the integration cannot be
interchanged. For a detailed discussion we refer to \cite{AKPW,Michael}.

\begin{raggedright}

\end{raggedright}

\end{document}